
\% PHYZZX TeX package
\input topp.tex

\PHYSREV
\tolerance 2000
\nopubblock
\titlepage

\title{FLUCTUATIONS OF THE INSTANTANEOUS SPACE AVERAGES IN
FRAGMENTATION WITH KOLMOGOROV TIME SCALES.}

\author{Sergei E. Esipov}

\address{James Franck Institute and Department of Physics, University of
Chicago, 5640 South Ellis Avenue, Chicago, Illinois 60637, USA}
\andaddress{ Department of Physics and Material Research Laboratory,
University of Illinois at Urbana-Champaign,
1110 West Green Street, Urbana, Il. 61801-3090, USA}
\vfil
\eject

\abstract{We investigate the time behavior of the
fragmentation model with Kolmogorov time scales and space contraction
resembling the random $\beta$-model.
The space averages computed at any instant using the entire
spatial realization of the
velocity field are shown to fluctuate strongly in time,
and in this sense they are not observable. The time averaging
required to achieve the steady-state distributions in the case of constant
energy supply
are found.
The fluctuations of the
probability distribution functions (PDFs) have intermittency
properties different
from those of PDFs.

PACS Numbers: 47.25.Cg, 05.40.+j

\chapter{Introduction}

The random $\beta$-model [1] and random multiplier approach to intermittency
in turbulence [2,3 and review 4]
correspond to the so-called generation expansion in the
theory of cascades [5,6]. We remind to the reader that these (fragmentation)
models
replace the real space by tree-like
branching space, and time is allowed to enter the formulas implicitly through
the generation number.
Both approaches have been useful in analyzing the available
experimental data [7] and providing modern language for turbulence, that of the
fractional dimensions [8].
The replacement of time by the generation number is fully justified
for self-averaging cascades which develop nearly-Gaussian properties for
fluctuations of the probability distribution functions (PDFs) [9,10].
Meanwhile, the
time scales of fluid turbulence lead to accelerating cascades when some
branches may reach arbitrarily small length scales while others have not
evolved at all. This case has been termed ``shattering'' [11], and
the main characteristic feature of shattering is that PDFs are not
observable in a single fragmentation process, instead fluctuations dominate
[9,10].
This motivated us to develop a model which contains two fragmentation
processes evolving simultaneously. The first is the fragmentation of the
energy carried by eddies, and the second is the volume occupied by the eddies.
In this model time parametrizes the evolution of the
measure (velocity pulsations) along the tree leading to time-dependent
multifractals, if one uses this name for intermittency.
Self-averaging cascades have identical multifractal properties for all
fluctuations, while the shattering may lead to different properties as we
shall see below.

We stress from the very beginning that the connection of the statistics
of the developed model with that of real turbulence is purely phenomenological
(the same is true for all the existing models of this sort). Namely, the model
may
reproduce some experimentally known features of the velocity statistics.
We are not solving the turbulence problem, but studying an interesting
fragmentation model.
Nevetheless, it is natural to borrow the language of turbulence for our
purposes.
The formulation of the model is very simple and its statistical
properties lead to a number of instructive questions
about Kolmogorov similarity
hypotheses. Two of these questions will be discussed in this article.

(1) The mean energy dissipation, $\langle\epsilon\rangle$,
is the only nontrivial parameter of K41 [12].
However, $\epsilon$ clearly fluctuates in time and space, so that values
$\langle\epsilon^q\rangle^{1/q}$  can be identified. K62 [13]
assumes log-normal
statistics of $\epsilon$. The distribution is borrowed by Obukhov [14] from
Kolmogorov's article [15], on {\it fragmentation}. Fragmentation leads
to log-normal law only if the time scale of each successive break-up is
the same (say, Yule-Furry cascade [6]).
This is clearly not the case for eddy fragmentation. In fact, for
accelerating cascades (which lead to shattering),
the distribution function of $\epsilon$ loses
any sense and is not observable.  One has to modify the similarity hypotheses
under these circumstances. This explains our interest to fragmentation
models [16].

(2) In fragmentation along the tree  the
space-averaging and time-averaging are quite different. In this sense the
Taylor hypothesis is invalid. Space averaging up to the scales of the
external scale $L$ is insufficient since the existing eddies of all scales are
generated by their common forefather. The time and conditions when this
parent eddy was split are already random, together with the entire space
statistics
at a given moment. Time-averaging over a sequence of largest eddies generated
by external "stirring" leads to proper averaging. One has to determine the
duration of the time-averaging needed.

\chapter{The fragmentation model}

Consider a container of liquid having the external length scale $L$. At time
zero the liquid is strongly agitated at length scales of order $L$, and
the resulted turbulent motion cascades to smaller scales and decays in time.
The present model assumes that we deal with fragmenting eddies and each
eddy {\it independently} decides when to split and what are the daugther
eddy sizes and energy contents.
When eddy splitting occurs the kinetic energy of the parent eddy $x_0$
is separated between (say) two daugther eddies in the amounts of $x_1$ and
$x_2 = x_0-x_1$. The splitting is governed by the following kernel
$$K(y,x) = x^{-\gamma} g(y/x),\eqn\kernel$$ which provides the
lifetime of the parent eddy to be $$\tau(x)^{-1} = \int_0^x dy
K(y,x) = c x^{-\gamma+1}.\eqn\twoeleven$$
Here
$c=\int_0^1 dz g(z)$ is the rate of fragmentation,
$g(z)=g(1-z)$ due to symmetry. The choice $\gamma={13\over{11}}$ corresponds to
Kolmogorov time scales. Indeed, if the velocity scales as $u\sim{l^
{1/3}}$, the characteristic time is $\tau\sim{l/u}\sim{l^{2/3}}$ [17].
The energy of the velocity pulsation is $x\sim{u^2l^3}\sim{l^{11/3}}$,
combining this with the characteristic time we get $\tau(x) = c^{-1}x^{2/11}$.
It is important for what follows that the power-law in \kernel, the value of
$\gamma$ is greater than 1.
In this case we deal with the cascade of
shattering type, which
leads to strong fluctuations [10].
The time of the entire cascade results from
$dx/dt=-x/\tau(x)$ and is of order of $\tau_0 \sim  c^{-1} x_0^{2/11}$.
We thus see that energy decay in the present model is exponential. This unreal
property disappears if we allow for a suitable
time-dependence of the rate $c$ (which then leads to power-laws).
In this paper we restrict ourselves
with constant (in space and time) rate $c$,
so that the model is applicable up to the largest turn-over time. The reason is
that we are interested here in the steady-state properties and later
introduce a constant external energy supply. In general, the
time and space-dependence of $c$ should
depend upon the course of fragmentation
and lead to non-linear feedbacks. The constant value of $c$ is a
mean-field-type simplification, which makes our cascade linear. The attempt to
connect $c$ and the energy dissipation rate does not seem quite
straightforward.

We would like to comment also on the influence of the viscosity.
It is understood that the change of the energy dissipation at small
scales may provide a
second bottleneck in the linear model and compete with the largest eddies.
The present model ignores any transport in the opposite direction, i.e.
aggregation, since the analysis of biased diffusion of the interface in the
tree-like space is not yet available.
Sufficiently strong aggregation processes may inhibit fluctuations.

The next stage is to specify the space contraction. Suppose, that the fraction
$\beta$ of the parent volume occupied by a daugther eddy is randomly
distributed within the region $0\leq\beta\leq 1$ with the probability
$P(\beta)$.
Suppose also that there is {\it
no} correlations between different realizations of
$\beta$ even if the eddies have a common parent. A given eddy of $n$-th
generation occupies volume $v_n = L^3\prod_{k=1}^{k=n}\beta_k$.
The energy conservation implies that the velocity of this eddy is of the
order of $u_n =
(2x_nL^3/v_n)^{1/2} \sim  2^{-n/2} \langle{\beta}^{-1/2}
\rangle_P^n$. The lenght scale of the eddy is $l_n =
v_n^{1/3} = L \langle{\beta}^{1/3}
\rangle_P^n$. Note that the energy dissipation does not enter any of these
relations and the model. It can be computed and investigated
in its own right.

\chapter{The evolution equation}

In the theory of fragmentation one introduces the probability functional
of having
a given {\it field}
of occupation numbers $N(x,v,t)$ of eddies of energy $x$ and volume $v$ at a
given time $t$,
$W\{N|x_0,v_0,t_0\}$, provided that the cascade starts with one
eddy of the energy $x_0$, volume $v_0 = L^3$ at time $t=t_0$ (cf. [10]).
The generating functional is,
$$G\{z|x_0,v_0,0\} = \int {\cal D}[N(x,v,t)] e^{\int dxdv z(x,v)N(x,v,t)}
W\{N|x_0,v_0,0\},\eqn\gene$$
and the backward evolution equation reads
$$\partial_tG\{z|x_0,v_0,0\} = - G\{z|x_0,v_0,0\}\int_0^{x_0} dy
K(y,x_0) + $$
$$\int_0^{x_0}dy K(y,x_0) \langle G\{z|y,\beta{v_0},0\}\rangle_P
\langle G\{z|x_0-y,\beta{v_0},0\}\rangle_P,\eqn\main$$
where the average is taken with the function $P(\beta)$. Eq\main\  can be
arrived at
as follows [10]. One starts with the backward Chapman-Kolmogorov equation [6]
for the particular initial condition of having one eddy at $x_0,v_0,t_0$, i.e.
with $W\{N|x_0,v_0,t_0\}$ and takes advantage of the tree-like structure
of the space. On the binary tree the break-up of the initial single eddy leads
to two eddies, each of them being described by the independent functional
$W\{N_i|x_i,v_i,t_0+dt\}$, $i=1,2$, $N_1+N_2 = N$, $x_1+x_2 = x_0$, $v_i=
\beta_i v_0$, here $\beta_{1,2}$ are two random values of $\beta$
generated for each of the eddies. The probability of having these two eddies
is just a product of the two functionals, it gives rise to the non-linear term
in \main. A given eddy is represented by a point in $\{x,v\}$-space,
so that the transition probability to the two daughter eddies is a
product of the kernel $K$ and function $P(\beta_1)P(\beta_2)$. We prefer
to write integration over daughter energies explicitly, while the
integration over possible values of $\beta_{1,2}$ is denoted by $\langle...
\rangle_P$.
Introducing moments
$$\langle{N(x,v,t|x_0,v_0,0)}\rangle_W
 = {\delta G\over{\delta z(x,v)}}_{|z=0},$$
$$\langle{N(x_1,v_1,t)N(x_2,v_2,t)|x_0,v_0,0}\rangle_W =
{\delta^2G\over{\delta z(x_1,v_1)\delta z(x_2,v_2)}}_{|z=0},\eqn\mom$$
we get closed equations for the moments. It is sufficient at this stage to
consider the quite simple model,
$$g(y,x) = c\delta(y-x/2), \quad P(\beta) = p\delta(\beta-b)+(1-p)
\delta(\beta-b^2),\eqn\choice$$
which leads to the equations
$$\tau_{\mu}\partial_t M_1(n,m|\mu,\nu) = \hat{\cal L} M_1(n,m|\mu,\nu)
+ \tau_{\mu}\delta_{n,\mu}\delta_{m,
\nu}\delta(t),\eqn\firstmom$$
$$\tau_{\mu}\partial_t M_2(n_1,m_1;n_2,m_2|\mu,\nu) = \hat{\cal L}
M_2(n_1,m_1;n_2,m_2|\mu,\nu)$$
$$ + 2\hat{\cal R}M_1(n_1,m_1|\mu,\nu)\hat{\cal R}M_1(n_2,m_2|\mu,\nu)
+\tau_{\mu}
\delta_{n_1,n_2}\delta_{n_1,\mu}\delta_{m_1,m_2}\delta_{m_1,\nu}\delta(t)
\eqn\twomom$$
with the operators
$$\hat{\cal R} M(...|\mu,\nu) = pM(...|\mu+1,\nu) + (1-p)M(...|\mu+1,\nu+1),
\eqn\operr$$
$$\hat{\cal L} M(...|\mu,\nu) = - M(...|\mu,\nu) + 2\hat{\cal R}
M(...|\mu,\nu),
\eqn\operl$$
and definitions
$$M_1(n,m|\mu,\nu) = \langle{N({x_0\over{2^n}},b^{n+m}v_0,t|
{x_0\over{2^{\mu}}},b^{\mu+\nu}v_0,0)}\rangle_W,$$
$$M_2(n_1,m_1;n_2,m_2|\mu,\nu) = \langle{N({x_0\over{2^{n_1}}},b^{n_1+m_1
}v_0,t|{x_0\over{2^{\mu}}},b^{\mu+\nu}v_0,0)}\times $$
$${N({x_0\over{2^{n_2}}}
,b^{n_2+m_2}v_0,t|{x_0\over{2^{\mu}}},b^{\mu+\nu}v_0,0)
}\rangle_W,\eqn\defim$$
$$\tau_n = c^{-1} (x_0/2^n)^{\gamma-1}.$$

The solution of the system
\firstmom\ is a bit less cumbersome for the Laplace transform
in time
$\tilde{M}_1 = \int_0^{\infty}dt e^{-st} M_1.$
It is given by
$$\tilde{M}_1(n,m|\mu,\nu)
 = 2^{n-\mu}\tau_{n}C_{n-\mu}^{m-\nu}p^{m-\nu}(1-p)^{n-\mu-m+\nu}
\prod_{k=\mu}^{k=n}(s\tau_{k}+1)^{-1},\eqn\greenf$$
and represents the (Laplace-transformed)
Green function of the entire hierarchy \main, here $C_n^m = n!/m!/(n-m)!$.
It is understood in \greenf\ that
all the index integers are non-negative. Using \greenf, one can write
the two-point correlator
$$\tilde{M}_2(n_1,m_1;n_2,m_2|\mu,\nu) = \delta_{n_1,n_2}\delta_{m_1,m_2}
\tilde{M}_1(n_1,m_1|\mu,\nu) + $$
$$2 \sum_{\mu',\nu'}
\tilde{M}_1(\mu',\nu'|
\mu,\nu)\int_{-i\infty}^{i\infty}{ds'\over{2\pi{i}}}\hat{R}\tilde{M}_1(
n_1,m_1|\mu',\nu';s')\hat{R}\tilde{M}_1(n_2,m_2|\mu',\nu';s-s')\eqn\nnm,$$
with the integration contour in $s'$-plane passing between the poles
of the two multipliers of the integrand and convention of non-negative
summation
ranges.

Note in \greenf\ that the volume
contraction factorizes out, and the bi-modal $P(\beta)$ leads to binomial
distribution.
Clearly, the binomial distribution is not a significant feature of the
model and one rather expects a certain limiting form for large $n,m$.
The precision of the central limit theorem which replaces any iterated
distribution by the Gaussian one is insufficient, since later we may need
high-order moments calculated with the help of \greenf.
The limiting form of the volume distribution is $\sim \exp[-nf({m\over{n}}
)]$ with some function $f$ which has received attention in the description of
multifractals [18]. For our binomial model it is obtainable from
Stirling expansion, $f(\xi) = \xi\ln\xi + (1-\xi)\ln(1-\xi)$. This
model has flexibility due to possible variations of the parameters
$p$ and $b$, so that one can mimic [1] the experimentally observable
intermittency corrections.

\chapter{Fluctuations of the energy cascade}

Since the volume distribution plays a passive role, let us first concentrate
on the energy cascade. We sum over possible realizations of
volume contraction, and Eq\main\ becomes especially simple for the
one-point generating function
$$\tau_{\mu}\partial_tG(n|\mu) = - G(n|\mu) + G(n|\mu+1)^2,\eqn\simpmain$$
with the argument notations of \defim\ and initial conditions
$$G(n|\mu;t=0) = 1 + \delta_{n,\mu}(e^z-1).\eqn\init$$
This system can be solved recursively,
$$G(n|n) = 1-e^{-t/t_n}+e^{z-t/t_n},$$
$$G(n,n-1) = 1 + {(1-2e^z+e^{2z})(e^{-2t/\tau_n}-e^{-t/\tau_{n-1}})\over{
1-{2\tau_{n-1}\over{\tau_n}}}} + {(2-2e^z)(e^{-t/\tau_{n-1}}-e^{-t/\tau_n})
\over{1-{\tau_{n-1}\over{\tau_n}}}},\eqn\recur
$$
although these expessions quickly become too cumbersome. The solution for
$G(n|\mu)$ with $n-\mu\gg1$ can be arrived at by arguing that only the
amplitude of the moments is a function of $n-m$, while the time-dependence
quickly becomes universal. One can check this proposition by examining
the Laplace transforms of the moments.
The nesting structure of Eq\simpmain\ results in
connections
$$\langle N^q(n|\mu;t)\rangle = 2^{q(\gamma-2)} \langle N^q(n|\mu-1;2^
{1-\gamma}t)\rangle.\eqn\funmom$$
so that the generating function satisfies the functional equation
$$\partial_tF(z,t) = - F(z,t) + F^2(2^{\gamma-2}z,
2^{1-\gamma}t),\eqn\fungen$$
with $F=G(n|\mu)$ and $t$ measured in units of $\tau_{\mu}$. We have only been
able to get the solutions for individual moments in terms of sums of infinite
products analogous to those in \greenf, \nnm. The convergence of the
time dependence of the moments to individual asymptotic forms is shown in
Fig.1. The ratios of the cumulants of the type $c_a^{1/a}/c_b^{1/b}$ become
universal functions, the simplest of them being
$${c_1\over{c_2^{1/2}}} = {\langle N(n|\mu;t)\rangle\over{(\langle N^2
(n|\mu;t)\rangle - \langle N(n|\mu;t)\rangle^2)^{1/2}}}.\eqn\signo$$
It is shown in Fig.2a for different values of $\gamma$, $1<\gamma<2$. Fig.2b
displays the next ratio, $c_3^{1/3}/c_2^{1/2}$.

We remind to the reader that the statistical properties of fragmentation
depend strongly upon the parameter $\gamma$. If $\gamma<1$ the generation
function has the asymptotic form [10]
$$G(n|\mu) = \exp(z\langle N(n|\mu)\rangle),\eqn\gauss$$
i.e. becomes essentially Gaussian.
The signal-to-noise ratio (not reflected in \gauss)
increases as the square root of the occupation number.
The case $\gamma = 1$ is the threshold of shattering [9,11]. The
expectation $\langle N(n|\mu;t)\rangle$ has the log-normal distribution
(there are corrections [9,11]); and there is still a window in time
when the signal-to-noise ratio grows to
infinity with the generation number. When $1<\gamma<2$ (``soft'' shattering),
the generating function acquires the special
universal form ($\gamma$-dependent) satisfying \fungen\ and described above.
The signal-to-noise
ratio reaches the finite maximum at some time of order of $\tau_0$.
The value of the maximum decreases from $\infty$ to 0 as $\gamma$ spans
from 1 to 2.  Finally, at $\gamma>2$ the generating function has the
asymptotic form
$$G(n|\mu) = 1+(e^z-1)\langle N(n|\mu)\rangle,\eqn\largegamma$$
which describes very strong fluctuations: all the cumulants are
approximately equal to
$\langle N(n|\mu)\rangle$, which vanishes as $n-\mu$ grows,
and so does the signal-to-noise
ratio at any time.

A study not presented here indicates that
the abovementioned properties of fragmentation do {\it not} depend upon the
choice of the break-up kernel \choice\ and the presence of volume
contraction. These alternations modify the shape of the distribution
function, say, the curves in Fig.2 would look distorted.

\chapter{The steady-state regime}

At this stage we have to replace the similarity hypotheses with the
results of the fragmentation model and compute averages of a certain field.
We have seen that the entire space averaging
is insufficient to get rid of fluctuations and therefore
{\it time-averaging} is needed. That is why it is easier to study
the system with constant external stirring, and introduce the
(random or periodic) supply of largest eddies. The steady-state
values of the generating function is connected with the time-dependent
one [10,19]
$$\ln \bar{G} = {1\over{\bar{\tau}}}\int_0^{\infty}dt \ln G(t),\eqn\steadygen$$
where $\bar{\tau}$ is the mean time between adding the large eddies with
energy $x_0$ due to the external source. In real steady-state
turbulence the parameters $\bar{\tau}$ and
$x_0$ are connected, since $\bar{\tau}$ is the largest turn-over time
$\bar{\tau}\sim L/u_0 \sim L^{5/2} x_0^{-1/2}$, so that $x_0 \sim
\bar{\tau}^{-2} L^5$.

The steady-state values of the moments are
$$\bar{M}_1(n,m|\mu,\nu) = {1\over{\bar{\tau}}}\int_0^{\infty}dt
M_1(n,m|\mu,\nu;t),\eqn\stf$$
$$\bar{M}_2(n_1,m_1;n_2,m_2|\mu,\nu) = {1\over{\bar{\tau}}}
\int_0^{\infty}dt M_2
(n_1,m_1;n_2,m_2|\mu,\nu;t) - $$
$${1\over{\bar{\tau}}}\int_0^{\infty}dt M_1(n_1,m_1|\mu,\nu;t)
M_1(n_2,m_2|\mu,\nu;t) - \bar{M}_1(n_1,m_1|\mu,\nu)\bar{M}_1
(n_2,m_2|\mu,\nu).\eqn\sts$$
The first follows from \greenf\ at $s=0$,
$$\bar{M}_1(n,m|\mu,\nu)
 = 2^{n-\mu}\tau_nC_{n-\mu}^{m-\nu}p^{m-\nu}(1-p)^{n-\mu-m+\nu}
,\eqn\fst$$
and is just the binomial distribution.
To get the two-point correlator one first evaluates \nnm\ at $s=0$,
$${1\over{\bar{\tau}}}\int_0^{\infty}dt M_2(n_1,m_1;n_2,m_2|\mu,\nu;t)
= \delta_{n_1,n_2}\delta_{m_1,m_2}
\bar{M}_1(n_1,m_1|\mu,\nu) + $$
$$2 \sum_{\mu'=\mu,\nu'=\nu}^{\mu'=\bar{\mu},\nu'=\bar{\nu}}
\bar{M}_1(\mu',\nu'|
\mu,\nu)\hat{R}\bar{M}_1(
n_1,m_1|\mu',\nu')\hat{R}\bar{M}_1(n_2,m_2|\mu',\nu') {\cal H}(n_1,n_2
|\mu'+1)
,\eqn\stnnm,$$
$$\bar{\mu} = \min(n_2-1,n_1-1),\quad \bar{\nu} = \min(m_1,m_2,\mu'),$$
and then computes
$${1\over{\bar{\tau}}}\int_0^{\infty}dtM_1(n_1,m_1|\mu,\nu;t)
M_1(n_2,m_2|\mu,\nu;t)=$$
$$\bar{M}_1(n_1,m_1|\mu,\nu)\bar{M}_1
(n_2,m_2|\mu,\nu)
{\cal H}(n_1,n_2|\mu).\eqn\msq$$
Here we used the function
$${\cal H}(n_1,n_2|\mu) = \sum_{\lambda=\mu}^{\lambda=n_1}
{1\over{\tau_{\lambda}}}
\prod_{k_1=\mu}^{k_1=n_1,
k_1\neq\lambda}\Big(1-{\tau_{k_1}\over{\tau_{\lambda}}}
\Big)^{-1}\prod_{k_2=\mu}^{k_2=n_2}\Big(1+{\tau_{k_2}\over{
\tau_{\lambda}}}\Big)^{-1},\eqn\calh$$
which is symmetric with respect to the first two arguments and non-negative.
Formulas \sts-\calh\ explicitly define the two-point correlator.

The amplitude of the first and second
moments grows with $n$ as $2^{n(2-\gamma)}$, $2^{2n(2-\gamma)}$ respectively.
The signal-to-noise ratio
$c_1^2/c_2 \sim \tau_0/{\bar{\tau}} \sim 1$ means that
the time averaging has to exceed greatly the largest turn-over time, $\tau_0$.
However, this
result is model-dependent; in our previous article [10] we found that
in the case of uniform continuous fragmentation the
signal-to-noise ratio is much smaller, $c_1^2/c_2 \sim
(x_{\min}/x_0)(\tau_0/{\bar{\tau}}) \ll 1$, where $x_{\min}$ is the energy
of the smallest eddy in the inertial range. The time required for
averaging has to exceed greatly $\tau_0 Re^{11/4}$, where $Re = u_0 L/\nu$ is
the Reynolds number.
Thus, performing more active stirring we increase the kinetic energy
of the large eddies and invoke additional paths to the eddies of a given
energy. As a result the steady-state fluctuations only increase.

\chapter{The velocity moments and energy dissipation}

In this Section we compute the moments of the
velocity pulsations.
Let us fix the
number $n+m = n_0$ and from \defim\ it follows that the averaging
volume $v_0b^{n_0}$ is fixed. Consider only the eddies having this
volume, note that they are distributed in their generations, and the occupation
numbers at the steady-state result from \fst
$${\cal N}
(j) = \bar{M}_1(j,n_0-j|0,0) = 2^{j}\tau_j
C_{j}^{n_0-j} p^{n_0-j}(1-p)^{2j-n_0},
\eqn\dist$$
with the range $n_0/2\leq{j}\leq{n_0}$.
The velocity moments which sample the space distribution are
defined as
$$\langle u_{n_0}^q\rangle_s \sim \sum_{j=n_0/2}^{j=n_0} \Big({2x_0\over{2^j
v_0b^{n_0}}}
\Big)^{q/2} N(j,n_0-j|0,0;t)
,\eqn\defvel$$
and fluctuate in time together with $N(j,n_0-j|0,0;t)$. The sign $\sim$
reflects the fact that this expession is not normalized.
In the presence of external stirring
these quantities become well-defined. It is of interest to sample the
fluctuations in time and time-average the moments
$\langle\langle u_{n_0}^{q/r}\rangle_s^r\rangle_t$, where
indexes $s,t$ denote space and time-averaging. At present we do not have a
nice analytical (or numerical) method to compute these averages at large $r$.

Consider the case $r=1$, when the time-fluctuations are not important.
The velocity moments,
$$\langle\langle u_{n_0}^q\rangle_s\rangle_t = A_1
\sum_{j=n_0/2}^{j=n_0} \Big({2x_0\over{2^jv_0b^{n_0}}}
\Big)^{q/2} {\cal
N}(j)
,\eqn\velone$$
can be analyzed with the focus on the intermittency properties in space.
Here $A$ is the normalization constant, $A_1^{-1} = \langle\langle 1
\rangle_s\rangle_t$.
Eqs\dist, \velone\ together with $l_{n_0} = Lb^{n_0/3}$ define the dependence
of the moments upon the eddy length scale, $n_0$ being the paramenter.
We have applied Eqs\dist, \velone\ to fit the experimental curve
of intermittency corrections provided that the third moment has the
exponent $\zeta_3=1$; our results are shown in Fig.3. The
best fit values of the parameters are $p=0.793$, $b=0.728$. The asymptotic
form for $\zeta_q = 1.454 + 0.138 q$ at large positive $q$.

The next case is $r=2$. The time-averaged moments are
$$\langle\langle u_{n_0}^{q/2}\rangle_s^2\rangle_t = A_2
\sum_{j_1=n_0/2}^{j_1=n_0}\sum_{j_2=n_0/2}^{j_2=n_0}
\Big({4x_0^2\over{2^{j_1+j_2}v_0^2b^{2n_0}}}\Big)^{q/4} {\cal
NN}(j_1,j_2),\eqn\veltwo$$
with the two-point correlator
$${\cal NN}(j_1,j_2) = \bar{M}_2(j_1,n_0-j_1;j_2,n_0-j_2|0,0)\eqn\twopoint$$
obtained in the previous Section.
The ``intermittency'' corrections computed with the help of \veltwo\
are shown by the dashed line in Fig.3, this curve look different, since
we have sampled a mixture of space and time fluctuations.

In addition we studied the average
$$\langle\langle\epsilon\rangle_s\rangle_t = A_{\epsilon}
\sum_{j=n_0/2}^{j=n_0}
\epsilon(j) {\cal
N}(j),\quad
\epsilon(j) = \Big({2x_0\over{2^jv_0b^{n_0}}}\Big)^{3/2} (v_0b^{n_0})^{-1/3}
,\eqn\ep$$
which can be identified with the energy dissipation averaged over eddies
of a given size. Fig.4 shows the dependence of this quantity versus the
length scale $l_{n_0}$ of the eddy. One can see that for $n_0 > 20$,
$\ln(l_{n_0}) < -2$, the energy dissipation is truly scale-independent, since
its exponent is given by $\zeta_3-1=0$, and meets the requirements postulated
by Landau \& Lifshitz [17]. Note, that the steady-state distribution of
$\epsilon$, which is given by ${\cal N}(j)$, Eq\fst\ (see Fig.5)
has a maximum at intermediate
$j$, while $j$ itself is proportional to $\ln\epsilon(j)$. Thus, namely
$\epsilon(j)$ has the distribution resembling log-normal one
(near the maximum), but not
the quantities $N(j,n_0-j|0;t)$ nor their spatial sums.

\chapter{Conclusion}

We have presented a model of fragmentation with space contraction
which can serve as a
phenomenological model of developed turbulence. Due to the
shattering the space averages fluctuate strongly in time.
It would be curious to
study the properties of space and time averaging in real stirred fluid
to make a more elaborated test of the model.
This model can also be applied to
the shell models [20] on branching trees.
There also, despite the
exponentially growing number of analogous branches, the average
over all branches of a given generation should still fluctuate
in time. Another interesting project is to study quantitatively the case when
the fragmentation (i.e. the eddy break-up) is accompanied by ``aggregation''
which proceeds at a comparable rate so that the net flux is still biased to
smaller eddy sizes.
The above developed method is not
applicable in this case, yet we expect the fluctuations to be of the
{\it same} order of magnitude.

\ack {I am grateful to L. Kadanoff for the encouragement to write this
article, to D.L. Maslov for numerios helpful telephone
conversations, to L. Kadanoff and D. Lohse for reading the manuscript and
helpful suggestions, to D. Lohse, M. Mungan and J. Wang for
communications on intermittency
at the meetings of the Royal Turbulence Club at the University of
Chicago. This work was
supported in part by the Material Research Laboratories at the
University of Illinois at Urbana-Champaign and at the University of Chicago,
and in part by NSF Grant NSF-DMR-89-20538.
}
\chapter{References}

\item{1.} R. Benzi, G. Paladin, G. Parisi and A. Vulpiani, J. Phys. A: Math.
Gen. {\bf 17}, 3521, (1984);
R. Benzi, L. Biferale, G.Paladin, A. Vulpiani and M. Vergassola,
Phys. Rev. Lett. {\bf 67}, 2299, (1991)
\item{2.} E. A. Novikov, Prikl. Mat. Mech., {\bf 35}, 266, (1971);
A. B. Chhabra and K.R. Sreenivasan, Phys. Rev. Lett. {\bf 68},
2762, (1992);
R. Benzi, L. Biferale, A. Crisanti, G.Paladin, M. Vergassola and
A. Vulpiani, Physica D, {\bf 65}, 352, (1993)
\item{3.} L. Biferale, M. Blank and U. Frisch, cond-mat network preprint,
(1993)
\item{4.} M. Grossmann and D. Lohse, Physica A, {\bf 194}, 519, (1993);
Z. Phys. B {\bf 89}, 11, (1992)
\item{5.}  See books and references therein. A.T. Bharucha-Reid, ``Elements of
the Theory of Markov Processes and Their Applications'', McGraw-Hill, New York,
1960; T.Harris, ``The Theory of
Branching Processes'', Springer-Verlag, Berlin, 1963
\item {6.} D.R. Cox and H.D. Miller,
``The Theory of Stochastic Processes'', Chapman and Hall, London, 1972;
S.K. Shrinivasan ``Stochastic
Theory and Cascade Processes'', American Elsevier, New York, 1969.
\item{7.} F. Anselmet, Y. Gagne,
E.J. Hopfinger and R.A. Antonia, J. Fluid Mech., {\bf 140}, 63, (1984)
\item{8.} C. Meneveau and K.R. Sreenivasan, Phys. Rev. Lett. {\bf 59}, 1424,
(1987); C. Meneveau and K.R. Sreenivasan,J. Fluid Mech., {\bf 224}, 429,
(1991);
M. Nelkin, J. Stat. Phys., {\bf 54}, 1, (1989); A.B. Chhabra, C. Meneveau,
R.V. Jensen and K.R. Sreenivasan, Phys. Rev. A, {\bf 40}, 5284, (1990);
A.B. Chhabra and K.R. Sreenivasan, Phys. Rev. A, {\bf 43}, 1114, (1991)
\item{9.} A.F. Filippov, Theory Probab. Its Appl. (USSR), {\bf 6},
275, (1961)
\item{10.} S.E.Esipov, L.P. Gor'kov and T.J. Newman, J.Phys.A: Math.
Gen. {\bf 26}, 787, (1993); D.L. Maslov, Phys. Rev. Lett., {\bf 71}, 1268,
(1993)
\item{11.} E.D. McGrady and R.M. Ziff, Phys.\ Rev.\ Lett.\ {\bf 58}, 892,
(1987)
\item{12.} A.N. Kolmogorov, Dokl. Akad. Nauk SSSR, {\bf 30}, 301, (1941);
{\bf 31}, 538, (1941); Proc. R. Soc. Lond. A, {\bf 434}, 9, 1991;
{\it ibid} {\bf 434}, 15, (1991); {\it ibid} {\bf 434}, 214, (1991)
\item{13.} A. M. Obukhov, J. Fluid Mech., {\bf 13}, 77, (1962)
\item{14.} A.N. Kolmogorov, J. Fluid Mech., {\bf 13}, 82, (1962)
\item{15.} A.N. Kolmogorov, Dokl. Akad. Nauk SSSR, {\bf 31}, 99, (1941)
\item{16.} We stress that the averaging over a number of
independent containers of fluid (which is sometimes used to arrive at
Taylor hypothesis)
has nothing to do with the space averaging inside a
single container.
\item{17.} L.D. Landau and E.M. Lifshitz, {\it Fluid Mechanics}, Pergamon
Press, NY, 1975
\item{18.} T.C. Halsey, M.H. Jensen, L.P. Kadanoff, I. Procaccia and
B.I. Shraiman, Phys.Rev.A, {\bf 33}, 1141, (1986)
\item{19.} I. Pazsit, J. Phys. D: Appl.Phys., {\bf 20}, 151, (1987)
\item{20.} E.B. Gledzer, Sov. Phys. Dokl. {\bf 18}, 216, (1973);
M. Yamada and K. Ohkitani, Prog. Theor. Phys., {\bf 81}, 329, (1989);
M.H. Jensen, G. Paladin and A. Vulpiani, Phys. Rev. A, {\bf 43}, 798,
(1991)

\chapter{Figure Captions}

\item{Fig.1} Convergence of the time-dependence of the moments $M_{\alpha}
(20|m;t) = \langle N^{\alpha}({x_0\over{2^{20}}},{x_0\over{2^m}};t)\rangle$
to the asymptotic form. The curves
are scaled in accordance with \funmom:
$\ln[M_{\alpha}(20|m;{t\over{\tau_m}})] + \alpha{m} \ln(2^{2-\gamma}
)$, $0\leq{m}\leq 19$, $\alpha = 1,2,3$, $c=1$.

\item{Fig.2} a) Time-dependence of the signal-to-noise ratio \signo, $n=20$,
$\mu = 0$, asymptotic for large $n$. The values of the contraction
factor, $b$ are indicated.

\item{ }
b) Time dependence of the skewness-to-deviation ratio, $c_3^{1/3}/c_2^{1/2}$
for the same
values of the parameters. In the fragmentation of this type
negative skewness
indicates that the the fluctuations are moderate (see our paper [10]).

\item{Fig.3} The intermittency exponent $\zeta_q$ defined via
$\langle\langle u_{n_0}^q\rangle_s\rangle_t \propto \langle\langle u_{n_0}
\rangle_s\rangle_t^{\zeta_q}$ with respect to the changes of $n_0$,
computed from \velone. The dots are the
experimental data [7]. The parameters from top to bottom, (a)
$p=0.808, b=0.73$, (b) $p=0.770, b=0.725$, (c) $p=0.734, b=0.72$.
These curves are independent upon the choice of $x_0, v_0$. Three significant
figures of $\zeta_q$ are obtained
provided that $n_0 > 20$. The dased line is computed with the help of
Eq\veltwo
using $p=0.793, b=0.728$.

\item{Fig.4} The scale dependence of the average energy dissipation,
Eq\ep. Note that for large enough $n > 20$, the dissipation is
indeed constant. The parameters, $2\leq n_0 \leq 256, p=0.793. b=0.728$.
For comparison, the first moment of the velocity is shown,
$\langle\langle u_{n_0}\rangle_s\rangle_t $, which has the exponent
$\zeta_1 = 0.368$.

\item{Fig.5} The logarithm of the distribution function, ${\cal N}(j),
$ Eq\dist\ versus the values of $\ln\epsilon(j)$ for $n_0 = 128$.
Note that the wings of the
surface are model-sensitive and would look different for a continuous
fragmentation in $\beta$.
This curve can be regarded as the $f(\alpha)$
curve, used in multifractal language. The insert shows the surface
made by such curves for $n_0 = 8j, 4\leq{j}\leq16$. The curves may be
superimposed, however, in this paper
we avoided using the saddle point approximation
in the continuous limit, since its precision is insufficient at large $q$.

\end